\documentstyle[aps,prb,preprint]{revtex}
\tightenlines
\begin{document}

\title{Density Functional Study of adsorption of molecular hydrogen on
graphene layers}
\author{J. S. Arellano \footnote{On sabbatical leave from Area de
F\'{\i}sica, Divisi\'{o}n de Ciencias B\'{a}sicas e Ingenier\'{\i}a,
Universidad Aut\'{o}noma Metropolitana Azcapotzalco, Av. San Pablo 180,
02200 M\'{e}xico D.F., M\'{e}xico.}, L.M. Molina, A. Rubio and J.A.
Alonso} \address{Departamento de F\'{\i}sica Te\'{o}rica,
         Universidad de Valladolid, 47011 Valladolid, Spain.}
\date{\today}
\maketitle


\begin{abstract} Density functional theory has been used to study
the adsorption of molecular $H_2$ on a graphene layer.  Different
adsorption sites on top of atoms, bonds and the center of carbon
hexagons have been considered and compared.  We conclude that the
most stable configuration of $H_2$ is physisorbed above the center
of an hexagon.  Barriers for classical diffusion are, however, very
small. \end{abstract}

\section {INTRODUCTION} \label{sec1}

The adsorption of hydrogen by different forms of carbon has been
studied by different groups. 
~\cite{Dillon,Darkrim,Chambers,Wang,Hynek,Ye,Liu} Dillon {\it et
al} ~\cite{Dillon} were the first to study the storage of molecular
hydrogen by assemblies of single wall carbon nanotubes (SWCNT) and
porous activated carbon.  They pointed out that the attractive
potential of the walls of the pores makes it possible a high
density storage. From temperature-programmed desorption experiments
Dillon {\it et al} ~\cite{Dillon} concluded that those forms of
carbon are promising candidates for hydrogen storage, although the
density of hydrogen is still low in order to meet the requirements
of the DOE Agency for novel hydrogen storage systems.  More
recently Levesque {\it et al} ~\cite{Darkrim}, Ye {\it et
al} ~\cite{Ye}, and Liu {\it et al} ~\cite{Liu} also studied the
adsorption of molecular hydrogen on SWCNT at different temperatures
and pressures. Chambers {\it et al} ~\cite{Chambers} have reported
obtaining an extraordinary storage capacity by some graphite
nanofibers but Wang and Johnson ~\cite{Wang} have tried
unsuccessfully to confirm the high storage capacity by graphite
nanofibers (slit pores) and SWCNT. Hynek {\it et al} ~\cite{Hynek}
investigated ten carbon sorbents but only one of them could augment
the capacity of compressed hydrogen gas storage vessels. The
improvement was marginal at 190 K and 300 K but nonexistent at 80
K. The storage capacity of carbon nanotubes and graphitic fibers
has been enhanced by doping with lithium and other alkali
elements~\cite{Chen}. The alkali atoms seem to have a catalytic
effect in dissociating the $H_2$ molecule and promoting atomic
adsorption. An advantage is that the doped systems can operate at
moderate temperatures and ambient pressure.

Some of the authors cited above ~\cite{Darkrim,Wang,Simonyan} have
also performed computer simulations of the adsorption of molecular
hydrogen inside, outside and in the interstices of an array of
SWCNT and in idealized carbon slit pores using model pair
potentials to describe the interactions. Wang and Johnson
~\cite{Wang} adopted the semiempirical pair potential of Silvera
and Goldman ~\cite{Silvera} for the $H_2-H_2$ interaction and the
$H_2-C$ interaction was modelled by a potential derived by Crowell
and Brown ~\cite{Crowell} by averaging an empirical Lennard-Jones
$H_2-C$ potential over a graphite plane. In the simulations Wang
and Johnson used a hybrid path integral-Monte Carlo method. Johnson
~\cite{Simonyan} also studied the influence of electrical charging
of the tubes.  Stan and Cole ~\cite{Stan} performed calculations
based on a sum of isotropic Lennard-Jones interactions between the
molecule and the C atoms of the tube.  They calculated the
adsorption potential of a hydrogen molecule, considered as a
spherically symmetric entity, as a function of distance from the
axis of a SWCNT, along radial lines upon the center of an hexagon
of carbon atoms and upon a carbon atom respectively.  Those
simulations give useful insight to interpret the results of the
experiments.  However the description of the interaction between
$H_2$ and the graphitic surfaces of the SWCNT or the slit pores in
those works is too simple.  Simplicity is a necessary requirement
for massive simulations involving several hundred (or several
thousand) $H_2$ molecules and an assembly of SWCNT of realistic
size, but one can expect more realistic results if the interaction
potential is derived from an ab initio calculation. The adsorption
of "atomic" hydrogen on a planar graphene sheet, that is a planar
layer exfoliated from graphite, has been studied previously
~\cite{Bercu,Jeloaica}. Bercu and Grecu ~\cite{Bercu} used a
semiempirical molecular orbital LCAO treatment at the INDO
(intermediate neglect of differential overlap) unrestricted
Hartree-Fock level and Jeloaica and Sidis ~\cite{Jeloaica} used the
Density Functional formalism (DFT)~\cite{Kohn-Sham}. In both works
the description of the graphene layers was simplified by modelling
this layer by a finite cluster C$_{24}$-H$_{12}$, where the
hydrogen atoms saturate the dangling bonds on the periphery of the
planar cluster. But, as mentioned above, hydrogen is adsorbed in
molecular form by graphitic surfaces (SWCNT and slit pores), so in
this work we study the interaction of an $H_2$ molecule with a
planar graphene layer.  Since the graphene layers interact weakly
in bulk graphite, the interaction of $H_2$ with a graphitic surface
is a localized phenomenon restricted to the outermost plane. For
this reason our calculations have relevance for understanding the
adsorption of $H_2$ on the walls of slit pores in graphite, and
also for the case of adsorption by SWCNT, since these differ from a
graphene layer only in the curvature of the layer.

\section {THEORETICAL METHOD AND TESTs} \label{sec2}

To calculate the interaction between $H_2$ and a planar graphene
layer we use the ab initio fhi96md code, developed by Scheffler
{\it et al} ~\cite{Bockstedte}. This code uses the DFT
~\cite{Kohn-Sham} to compute the electronic density and the total
energy of the system, and we have chosen the local density
approximation (LDA) for exchange and correlation.~\cite{Perdew}
Only the two electrons of the $H_2$ molecule and the four external
electrons ($2s^2 \, 2p^2$) of each carbon atom are explicitly
included in the calculations, while the $1s^2$ core of carbon is
replaced by a pseudopotential. For this purpose we use the nonlocal
norm-conserving pseudopotentials of Hamann {\it et al}
~\cite{Bachelet,Hamann}. Nonlocality in the pseudopotential is
restricted to $\ell = 2$, and we take as local part of the
pseudopotential the $s$ component.  The code employs a supercell
geometry and a basis of plane waves to expand the electronic wave
functions.~\cite{Payne}

First we have tested the method for pure graphite. By minimization
of the total energy with respect to the interatomic distances we
obtained an in-plane $C-C$ bond length equal to 2.66 a.u. and a
distance between planar graphitic layers of 6.275 a.u. The
corresponding experimental values ~\cite{Dresselhaus} are 2.68 a.u.
and 6.34 a.u. respectively. The small ($1 \%$) underestimation of
bond lengths is characteristic of the LDA.  Next we have studied an
isolated graphene layer. Since the computer code uses a periodic
supercell method, the cell axis has to be large in the z-direction
to avoid the interaction between graphene sheets in different
cells.  Table~I gives the calculated energy of the graphene layer
as a function of the length c of the unit cell in the z direction,
or in other words, as a function of the distance between parallel
graphene layers.  Results given for c = 20, 25, 30 and 35 a.u. show
that the energy is well converged for those layer separations and
that for c = 20 a.u. the error in the energy per atom is only about
1 in $10^5$. A cutoff of 40 Ry was used in all the calculations. 
We also tested the method by calculating the energy of the $H_2$
molecule, that was placed at the center of a simple cubic
supercell. The total energy obtained for a plane wave cut-off
energy of 40 Ry and supercell lattice constants of 18 a.u. and 20
a.u. is the same, -2.247 Ry as well as the bondlength, 1.48 a.u.
Notice that this bond length is small compared to the $C-C$ bond
length. Anticipating the geometry to be used in the study of the
interaction between $H_2$ and graphene, another set of calculations
were performed for the energy of $H_2$ by placing the molecule in
the superlattice described above in the study of the graphene
layer, but this time without graphene. Calculations for distances
between the imaginary graphene planes ranging from 20 a.u. to 35
a.u. (the plane wave cut-off was again 40 Ry) gave energies for the
$H_2$ molecule, identical to the energies obtained for the cubic
superlattice geometry. 

\section {INTERACTION BETWEEN $H_2$ AND THE GRAPHENE LAYER} 
\label{sec3}

For the periodicity of the system we have selected a unit cell with
eight carbon atoms and one hydrogen molecule (see Fig. 1). If we
place a hydrogen molecule at any point of the cell, the distance
from this molecule to other in the nearest cells is 9.224 a.u. This
separation is large compared to the bond length of $H_2$ (1.480
a.u.), and we have verified that there is no interaction between
two hydrogen molecules separated by that distance. The interaction
of the $H_2$ molecule with the graphene sheet has been studied by
performing static calculations for two orientations of the axis of
the molecule: axis perpendicular to the graphene plane and axis
parallel to that plane.  Three possibles configurations, called A,
B and C below, have been selected for the perpendicular approach of
the molecule to the plane:  (A) upon one carbon atom, (B) upon the
center of a carbon-carbon bond, and (C) upon the center of an
hexagon of carbon atoms. On the other hand, for the parallel
approach the molecule is placed upon the center of an hexagon of
carbon atoms with the molecular axis perpendicular to two parallel
sides of the hexagon, and this is called configuration D. These four
configurations are given in the bottom panel of Fig. 1. To obtain
the interaction energy curve for each of those four cases, the
distance between the hydrogen molecule and the graphene layer was
varied while maintaining the relative configuration. In these
calculations the bond length of the $H_2$ was held fixed at 1.48
a.u., the bondlength of the free molecule. This is expected to be
valid in the relevant region of the interaction. This constraint
will, however, be relaxed in simulations described at the end of
this section.  Calculations were first performed in the parallel
configuration (D) for a superlattice such that the distance between
graphene layers is 30 a.u. The plane wave cut-off was 40 Ry. The
interaction energy curve is plotted in Figure 2 and the curve has a
minimum at 5.07 a.u. For separations larger than this value the
energy rises fast and reaches its asymptotic value for 10 - 11 a.u.
The energy at the maximum possible separation between the center of
mass of the $H_2$ molecule and the graphene plane for this
superlattice, 15 a.u., was taken as the zero of energy. The figure
also gives the results of a similar calculation for a smaller
superlattice, such that the distance between graphene layers is 20
a.u. The corresponding energy curve, referred to the same zero of
energy as above, is practically indistinguishable from the former
curve. The calculations also show that for all practical purposes
the energy curve has reached its asymptotic value for a distance of
10 a.u., that is the longest separation allowed for the
superlattice of 20 a.u. This indicates that calculations using the
smaller superlattice are enough for our purposes of studying the
$H_2$ - graphene interaction. Then, the results of calculations
corresponding to configurations A, B, C and D for a superlattice of
20 a.u. are given in Figure 3. The potential energy curves for the
perpendicular approach (A, B, C) rapidly merge with each other for
large $H_2$ - graphene separation, becoming indistinguishable from
one another beyond 6.5 a.u. Actually, curves A and B are very close
in the whole range of separations although B is marginally more
attractive. The common value of the energy of curves A, B and C at
separation 10 a.u. is taken as zero of energy in Figure 3. The
predicted equilibrium positions and the binding energies (depth of
the minimum) of the different curves are given in Table II. The
small magnitude of the binding energies, less than 0.1 eV, shows
that the system is in the physisorption regime.  Comparison of the
four curves reveals that the most favorable position for the $H_2$
molecule is physisorbed in a position above the center of a carbon
hexagon, and that the parallel configuration is slightly more
favorable than the perpendicular one. We have verified that
different orientations of the molecular axis with respect to the
underlying carbon hexagon in the parallel configuration lead, in
all cases, to the same curve D plotted in Fig. 3. The differences
in binding energy shown in Table II are very small. For instance,
configuration D and A only differ by 16 meV, and configuration D
and C by 3 meV. 

Figures 4a and 4b give the electron density of the pure graphene
layer in two parallel planes, 5 and 3 a.u.  above the plane of the
nuclei, respectively. The former one is very close to the preferred
distance of approach for the $H_2$ molecule in configuration D.
First of all one can note that the values of the electron density
in that plane are very small, of the order of $10^{-5}$ e/$({\rm
a.u.})^3$, so the plane is in the tail region of the electron
density distribution. Nevertheless the densities clearly reveal the
topography of the graphene layer. Electron density contours on top
of carbon atoms surround other contours representing the large
hexagonal holes.  Densities are larger in the other plane, closer
to the plane of nuclei. In each plane the density is larger in the
positions above carbon atoms and lower above the hexagons.  A plot
that complements this view is given in Figure 5, that gives the
electron density in a plane perpendicular to the graphene layer
through a line containing two adjacent carbon atoms, labelled $C_1$
and $C_2$ in the figure. Then, points labelled M and X represent
the midpoint of a carbon-carbon bond and the center of an hexagon
respectively. The most noticeable feature is the existence of
depressions of electron density in the regions above the centers of
carbon hexagons. These hollow regions are separated by regions of
larger density that delineate the skeleton of carbon-carbon bonds.
In this figure the density of the most external contour is $\rho =
 1.11 \times 10^{-2}$ e/$({\rm a.u.})^3$ and the interval between
contours $\Delta \rho = 1.11 \times 10^{-2}$ e/$({\rm a.u.})^3$

These observations correlate with the features in Fig. 3, and lead
to the following interpretation of the potential energy curves.
Each curve can be seen as arising from two main contributions, one
attractive and one repulsive.  The attractive contribution is
rather similar for all the configurations (notice the similarity of
the potential energy curves beyond 6 a.u.) and is mainly due to
exchange and correlation effects.  Neglecting correlation for the
purposes of simplicity, the exchange contribution to the total
energy is given, in the LDA, by the functional

\begin{equation} 
E^{LDA}_x[\rho] = C_x\int\!\rho({\bf r})^{4 \over3} $ $d^3 r  ,
\label{ec1} 
\end {equation} 
where $C_x$ is a well known negative constant.~\cite{Kohn-Sham} In
the regime of weakly overlapping densities, and assuming no density
rearrangements due to the close-shell character of $H_2$, the
contribution of exchange to the interaction energy becomes

\begin{equation}
\Delta E_x  = C_x [ \int[\rho_{H_2}{\bf (r)}\\ +
\rho_{g}{\bf (r)}\\]\,^{4 \over3}\,  d^3 r
-  \int\rho_{H_2}{\bf (r)}\\^{4 \over 3}\,  d^3 r
 -  \int\rho_{g}{\bf (r)}\\^{4 \over 3}\,  d^3 r ] ,
\label{ec2}
\end {equation}
where $\rho_g$ and $\rho_{H_2}$ represent the tail densities of the
graphene layer and $H_2$ molecule respectively.  A net "bonding"
contribution arises from the nonlinearity of the exchange energy
functional. On the other hand the sharp repulsive wall is due to
the short-range repulsion between the close electronic shell of the
$H_2$ molecule and the electron gas of the substrate. This
contribution is very sensitive to the local electron density
sampled by the $H_2$ molecule in its approach to the graphene layer
and explains the correlation between the position and depth of the
different minima in Fig. 3 and the features of the substrate
electron density in Figs. 4 and 5. Similar arguments explain the
physisorption of noble gas atoms on metallic surfaces~\cite{Lang}
and the weak bonding interaction between noble gases.~\cite{Gordon}
At very large separation the interaction energy curves should
approach the Van der Waals interaction, that is not well described,
however, by the LDA. 

An interesting point concerns the comparison of the minima of the
curves C and D of Fig. 3. That of curve D is deeper and occurs at a
shorter $H_2$ - graphene separation. The reason is that the
surfaces of constant electron density of the $H_2$ molecule have
the shape of slightly prolate ellipsoids instead of simple spheres. 
Consequently, for a given distance {\it d} between the center of
mass of $H_2$ and the graphene plane, the molecule with the
perpendicular orientation (C) penetrates more deeply into the
electronic cloud of the substrate than in the parallel orientation
(D). In other words, the repulsive wall is reached earlier, that is
for larger {\it d}, in the perpendicular configuration (C). If we
consider an electronic density contour in $H_2$ with a value $\rho
=0.018 $ e/$({\rm a.u.})^3$, then the two semiaxes have lengths of
2.07 and 1.71 a.u. respectively and the difference between these
two lengths is 0.36 a.u.  This value is in qualitative agrement
with the difference between the $H_2$~-~graphene~separations for
the two minima of curves C and D, which is 0.20 a.u. This shape
effect is usually neglected in the phenomenological approaches,
that treat $H_2$ simply as a spherical molecule. 

Figure 6 gives a plot of the charge density difference
\begin{equation}
\rho_{diff}{\bf(r)}\\= \rho_{tot} {\bf(r)}\\ -
(\rho_g{\bf(r)}\\
+  \rho_{H_2} {\bf(r)}\\),
\label{ec3}
\end {equation}
where $\rho_{tot}{\bf(r)}$ is the calculated density of the total
system, that is the $H_2$ molecule physisorbed in orientation D at
a distance of 5 a.u. above the graphene layer, whereas $\rho_g$ +
$\rho_{H_2}$ is the simple superposition of the densities of the
pure graphene layer and $H_2$ molecule 
placed also in orientation D, 5 a.u. above the graphene layer. That
density difference $\rho_{diff}{\bf(r)}\\$ is given in the same
plane, perpendicular to the graphene layer, used in Fig. 5.
$\rho_{diff}{\bf(r)}\\$ has positive and negative regions. The
positive region is the area bound by the contour labelled P. This
region has the shape of two lobes joined by a narrow neck. Contour
P has a value $\rho_{diff} = 2.36 \times 10^{-5}$ e/$({\rm
a.u.})^3$ and $\rho_{diff}$ increases in this positive region as we
move towards inner contours in the lobes. The innermost contour
shown has a value $\rho_{diff} = 2.87 \times 10^{-4} $ e/$({\rm
a.u.})^3.$ The $H_2$ molecule sits above the neck, so the figure
reveals that the repulsive interaction produced by the close
electronic shell of $H_2$ pushes some charge from the region
immediately below the molecule (the neck region) to form the lobes
of positive $\rho_{diff}{\bf(r)}.$ This displacement of electronic
charge is nevertheless quantitatively very small. Notice that
$\rho_g$ takes values between $ 1.6 \times 10^{-3} $ and $ 4.1
\times 10^{-3}$ e/$({\rm a.u.})^3$ in a plane 3 a.u.  above the
graphene layer, while $\rho_{diff}$ has values of the order
$10^{-5} - 10^{-4}$ e/$({\rm a.u.})^3$ in the same plane. The
smallness of $\rho_{diff}$ justifies the argument given in eq. (2)
for the attractive exchange-correlation contribution to the
interaction potential. 

The static calculations discussed above have been complemented with
dynamical simulations in which the $H_2$ molecule was initially
placed in different orientations at distances of 4 - 6 a.u. from
the graphene layer and was left to evolve under the influence of
the forces on the H atoms.  The $H_2$ bondlength was allowed to
adjust in the process. The simulations confirm the results of the
static calculations, in the sense that the $H_2$ molecules end up
in positions above the center of an hexagon at the end of the
simulations. The binding energies and $H_2$ - graphene layer
distances practically coincide with those in Table II. Marginally
small differences in separation or binding energy are due to very
small changes of the bondlength of $H_2$, always smaller than $0.3
\%$. The result of one of the simulations is worth to be mentioned.
A configuration intermediate between those labelled C and D above
was obtained: the center of mass of the molecule was 5.10 a.u.
above the center of a carbon hexagon, with the molecular axis
forming an angle of about $30^{\circ}$ with the graphene plane. The
binding energy in this new configuration was only 1 meV larger than
in the parallel configuration D. 

In summary, the picture arising from the calculations is rather clear. The
$H_2$ molecules prefer the hollow sites above the centers of carbon
hexagons where the background electron density is lower than in channels
on top of the skeleton of carbon-carbon bonds. The exchange-correlation
contribution provides the weak attraction responsible for physisorption,
but the preferred distance of approach is determined by the repulsive part
of the interaction potential. That repulsive contribution is due to the
close-shell electronic structure of $H_2$. We have performed static
calculations of the barrier for the diffusion of a molecule, initially in
the parallel configuration D at the preferred distance of 5.07 a.u. above
the graphene plane, to an equivalent configuration D above an adjacent
hexagon. The initial configuration of the molecule, with its axis
perpendicular to two parallel carbon-carbon bonds, can be seen in the
bottom panel of Fig. 1. The molecule was then forced to follow a path
across one of those bonds, allowing for the reorientation of the molecular
axis at each step in order to minimize the energy of the system. Although
the molecule begins with the axis parallel to the graphene plane, the
orientation of the axis changes as the molecule approaches the
carbon-carbon bond. In fact, when the center of mass of the molecule is
precisely above that bond, the molecular axis becomes perpendicular to the
graphene plane, that is the molecule adopts configuration B, as indicated
also in Fig. 1. The energy difference between this saddle configuration
and the starting one gives a calculated diffusional barrier of 14 meV. A
temperature of 163 K is enough to surpass this barrier. 

The conclusions from the calculations are, in our view, general
enough that one can make some extrapolations to the case of
adsorption of $H_2$ by carbon nanotubes. When adsorption occurs on
the outside wall of an isolated nanotube, the predictions of Fig. 3
will be valid, with a minor influence of the nanotube curvature. If
the tubes form a parallel bundle and we consider the interstitial
channels between tubes, the effects seen in Figure 3 will be
smoothed out because of the addition of contributions of different
graphitic surfaces non in registry. Addition of these contributions
will give rise to an interstitial channel with a potential energy
nearly independent of z, if we call z the direction parallel to the
tube axis.  Finally, the same smoothing effect will occur in the
inner channel of a tube if the tube diameter is not large. In
summary we predict very easy diffusion of the $H_2$ molecule in
arrangements of parallel tubes along the direction parallel to the
tube axis, both inside the tube cavity and in the interstitial
channels.  

  The present adsorption results can be partialy compared with
those of Stan and Cole.~\cite{Stan} They considered the $H_2$
molecule as a spherically symmetric entity and calculated the
adsorption potential inside zigzag (13,0) nanotubes (radius $ = $
9.62 a.u.)  based on a sum of isotropic Lennard-Jones interactions
between the molecule and the carbon atoms of the tube. Our
calculation and that of Stan and Cole agree in that the smallest
binding energy is obtained for the $H_2$ upon one carbon atom and
the largest one for the $H_2$ upon the center of the hexagon of
carbon atoms. However Stan and Cole do not distinguish between
parallel and perpendicular orientations because they considered an
spherical molecule. Their Fig. 1 shows a binding energy about 0.079
eV for adsorption in front of the center of an hexagon of carbon
atoms and that the equilibrium distance between the molecule and
the nanotube wall is 5.7 a.u. This distance is consistent but a
little larger than those reported in our Table II.  On the other
hand, the value 0.079 eV for the binding energy is also consistent
with the binding energies in Table II. Notice, however, that the
binding energy for a tube of larger radius, or for a planar
graphene sheet, will be a little smaller because the curvature of
the tube increases the number of nearest neighbor carbon atoms. In
fact, Wang and Johnson ~\cite{Wang} calculated an adsorption
binding energy near 0.050 eV for molecular hydrogen in an idealized
carbon slit pore with a pore width of 17.4 a.u. 

\section{CONCLUSIONS}  

By performing DFT calculations we confirm that physisorption of
$H_2$ on graphitic layers is possible. The differences between the
binding energies corresponding to different positions (on top of
carbon atoms, on top of carbon-carbon bonds, on top of hexagonal
holes) are small, and the diffusional barriers are also small, so
easy diffusion is expected at low temperature. The nonsphericity of
the $H_2$ molecule has some influence on the preferred orientation
of the molecular axis with respect to the graphene plane. These
small effects associated to different positions and orientations of
the physisorbed molecule are expected to average out inside carbon
nanotubes or in the interstitial channels in parallel arrays of
carbon nanotubes. 

\section*{acknowledgements}

Work supported by DGES(Grant PB95-0720-C02-01), Junta de
Castilla y Le\'on (Grant VA28/99) and European Community (TMR
Contract ERBFMRX-CT96-0062-DG12-MIHT). L.M.M. is greatful to
DGES for a Predoctoral Grant.  J.S.A. wishes to thank the
hospitality of Universidad de Valladolid during his
sabbatical leave and grants given by Universidad Aut\'onoma
Metropolitana Azcapotzalco and by Instituto Polit\'ecnico
Nacional (M\'exico). Finally we thank the referee for
constructive suggestions.


\begin{table}
\caption {Calculated energy (Ry) of the graphene layer for several
layer-layer distances. The energies are calculated for a plane wave
cut-off energy of 40 Ry.}
\begin {tabular}{ccc}\hline
Layer-layer distance (a.u.) & Energy (per atom) \\
\hline 
       20 & -11.4234 \\
       25 & -11.4235 \\
       30 & -11.4235 \\
       35 & -11.4235 \\
\end{tabular}
\label{table1}
\end{table}

\begin{table}
\caption{Binding energy (eV) and equilibrium distance (a.u.) for $H_2$
physisorbed on a graphene layer. A, B and C correspond to the
configurations in which the molecular axis is perpendicular to the
graphene plane and the molecule is on top of: a carbon atom (A), the
midpoint of a carbon-carbon bond (B), the center of an hexagon (C). In
configuration D the molecule is on top of the center of an hexagon with
the molecular axis parallel to the graphene plane.}
\begin {tabular}{ccccc}\hline 
 & A & B & C & D \\
\hline
Binding energy  & 0.070 & 0.072 & 0.083 & 0.086 \\
Distance   & 5.50 & 5.49 & 5.25 & 5.07 \\
\end{tabular} 
\label{table2}
\end{table}
\begin{figure}
\caption{Top panel gives a fragment of the graphene layer showing
the eight carbon atoms in the unit cell, represented by large
spheres. Bottom panel shows the three adsorption configurations
with the molecular axis perpendicular to the graphene plane.  These
have the $H_2$ molecule above a carbon atom (A), above the midpoint
of a carbon-carbon bond (B) and above the center of an hexagon (C).
Also shown is an adsorption configuration (D) with the molecular
axis parallel to the graphene plane and the molecule above the
center of an hexagon.}
\label{fig1}
\end{figure}  
\begin{figure}
\caption{Comparison of potential energy curves for the parallel
approach of $H_2$ to the graphene layer upon the center of an hexagon
of carbon atoms. The curves were obtained using supercells such that
the graphene layers are separated by 30 a.u. (circles) or 20 a.u.
(crosses).}
\label{fig2}
\end{figure}
\begin{figure}
\caption{Potential energy curves for the aproach of $H_2$ to the
graphene layer in four different configurations. The axis of the
molecule is perpendicular (A, B, C) or parallel (D) to the graphene
layer. In the former orientation the molecule is above a carbon
atom (A), above the center of a C-C bond (B), and above the center
of an hexagon (C). In the parallel orientation (D) the molecule is
above the center of an hexagon.}
\label{fig3}
\end{figure}
\begin{figure}
\caption{(a) Contours of constant electron density $\rho$ of a pure
graphene layer in a plane 5 a.u. above the plane of the carbon
nuclei. $\rho = 8.64 \times 10^{-5}$ e/$({\rm a.u.})^3$ in the
innermost contours above carbon atoms. $\rho = 6.08 \times 10^{-5}$
e/$({\rm a.u.})^3$ in the innermost contours above the large
hexagonal holes.  Densities decrease monotonously between those two
contours with an interval $\Delta\rho = 0.18 \times 10^{-5}$
e/$({\rm a.u.})^3$. (b)  Contours in a plane 3 a.u. above the plane
of carbon nuclei. $\rho = 3.92 \times 10^{-3}$ e/$({\rm a.u.})^3$
in the innermost contours above carbon atoms. $\rho = 1.72 \times
10^{-3}$ e/$({\rm a.u.})^3$ in the innermost contours above the
large hexagonal holes. Densities decrease monotonously between
those two contours with an interval $\Delta \rho = 0.16 \times
10^{-3}$ e/$({\rm a.u.})^3$}
\label{fig4} 
\end{figure}
\begin{figure} 
\caption {Contours of constant electron density of pure graphene in
a plane perpendicular to the graphene layer, going through a line
containing two adjacent carbon atoms, labelled $C_1$ and $C_2$. 
Symbols M and X indicate the mid-point of a carbon-carbon bond and
the center of an hexagon, respectively. The outermost contour
plotted is $\rho = 1.11 \times 10^{-2}$ e/$({\rm a.u.})^3$ and the
interval between contours $\Delta \rho = 1.11 \times 10^{-2}$
e/$({\rm a.u.})^3$}
\label{fig5} 
\end{figure}
\begin{figure} 
\caption{Charge density difference $\rho_{diff} = \rho_{tot} -
\rho_g - \rho_{H_2}$ for $H_2$ physisorbed 5 a.u. above the
graphene layer. The plane of the plot and the symbols C1, C2, M and
X are the same as in Fig. 5. Contour labelled P has a value $2.36
\times 10^{-5}$ e/$({\rm a.u.})^3$ and encloses the region of
positive $\rho_{diff}.$}
\label{fig6} 
\end{figure} 

\begin{thebibliography}{99}
\bibitem{Dillon} A.C. Dillon, K.M. Jones, T.A. Bekkedahl, C.H. Kiang, D.S.
Bethune and M.J. Heben, Nature (London) {\bf 386}, 377 (1997).
\bibitem{Darkrim} F. Darkrim and D. Levesque, J. Chem. Phys. {\bf 109},
4981 (1998);  F. Darkrim, J. Vermesse, P. Malbrunot and D. Levesque, J.
Chem. Phys.  {\bf 110}, 4020 (1999).
\bibitem{Chambers} A. Chambers, C. Park, R.T.K. Baker and N.M. Rodriguez,
J. Phys. Chem. {\bf 102}, 4253 (1998).
\bibitem{Wang} Q. Wang and J. K. Johnson, J. Chem.  Phys. {\bf 110}, 577
(1999); J. Phys. Chem. B {\bf 103}, 4809 (1999).
\bibitem{Hynek} S. Hynek, W. Fuller and J. Bentley, Int. J. Hydrogen
Energy {\bf 22}, 601 (1997).
\bibitem{Ye} Y. Ye, C.C. Ahn, C. Witham, B. Fultz, J. Liu, A.G. Rinzler,
D. Colbert, K.A. Smith and R.E. Smalley, Appl. Phys. Lett. {\bf 74}, 2307
(1999).
\bibitem{Liu} C. Liu, Y.Y. Fan, M. Liu, H.T. Cong, H.M. Cheng and M.S.
Dresselhaus, Science  {\bf 286}, 1127 (1999).
\bibitem{Chen} P. Chen, X. Wu, J. Lin and K.L. Tan, Science {\bf 285}, 91
(1999).
\bibitem{Simonyan} V.V. Simonyan, P. Diep and J.K. Johnson, J. Chem. Phys.
{\bf 111}, 9778 (1999).
\bibitem{Silvera} I.F. Silvera and V.V. Goldman, J. Chem. Phys. {\bf 69},
4209 (1978).
\bibitem{Crowell} A.D. Crowell and J.S. Brown, Surf. Sci. {\bf 123}, 296
(1982).
\bibitem{Stan} G. Stan and M.W. Cole, J. Low Temp. Phys.  {\bf 110}, 539
(1998).
\bibitem{Bercu} M.I. Bercu, V.V. Grecu, Romanian J. of Phys. {\bf 41}, 371
(1996).
\bibitem{Jeloaica} L. Jeloaica and V. Sidis. Chem. Phys. Lett. {\bf 300},
157 (1999).
\bibitem{Kohn-Sham} W. Kohn and L.J. Sham, Phys. Rev {\bf 140}, A1133
(1965);
R.G. Parr and W. Yang, Density Functional Theory of Atoms and Molecules,
Oxford University Press, New York (1989).
\bibitem{Bockstedte} M. Bockstedte, A. Kley, J. Neugebauer and M.
Scheffler. Comp. Phys. Commun. {\bf 107}, 187 (1997).
\bibitem{Perdew} J.P. Perdew and A. Zunger, Phys. Rev. B {\bf 23}, 5048
(1981).
\bibitem{Bachelet} G.B. Bachelet, D.R. Hamann, and M. Schluter, Phys.
Rev. B {\bf 26}, 4199 (1982).
\bibitem{Hamann} D.R. Hamann, Phys. Rev. B {\bf 40}, 2980 (1989).
\bibitem{Payne} M.C. Payne, M.P. Teter, D.C. Allan, T.A. Arias and J.D.
Joannopoulos, Rev. Mod. Phys. {\bf 64}, 1045 (1992).
\bibitem{Dresselhaus} M.S. Dresselhaus, G. Dresselhaus and P.C. Eklund,
Science of Fullerenes and Carbon Nanotubes, Academic Press, San Diego
(1996).
\bibitem{Lang} N.D. Lang, Phys. Rev. Lett. {\bf 46}, 482 (1981).
\bibitem{Gordon} R.G. Gordon and V.S. Kim, J. Chem. Phys. {\bf 56}, 3122
(1972).
\end{thebibliography}
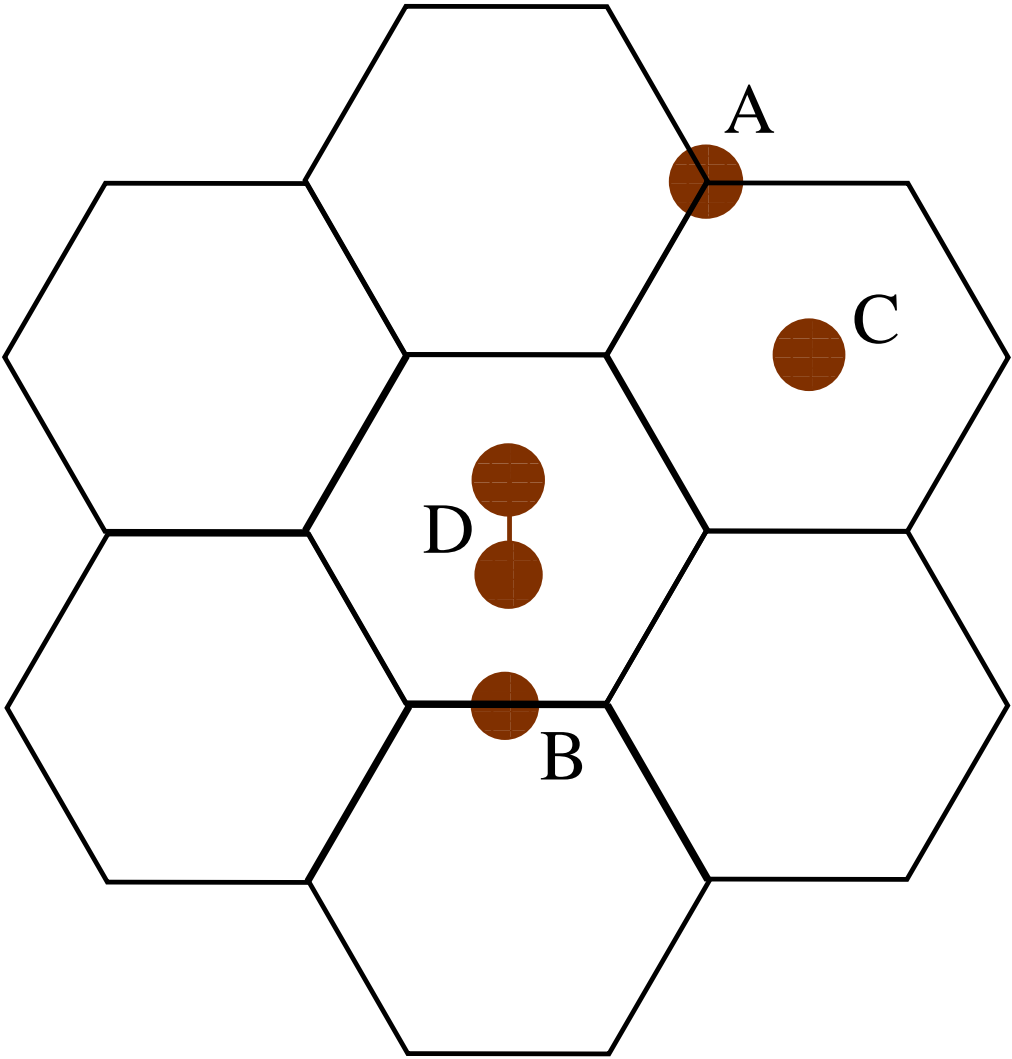
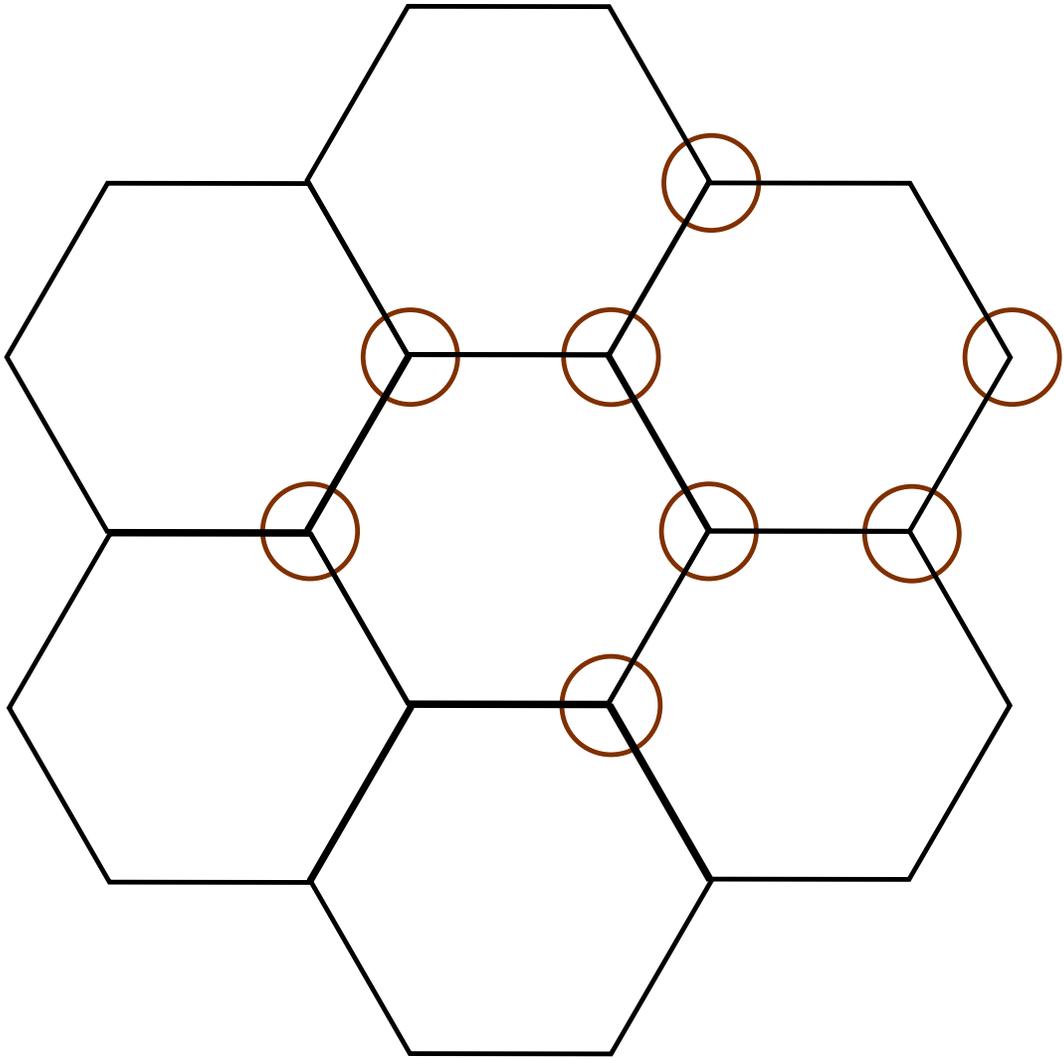
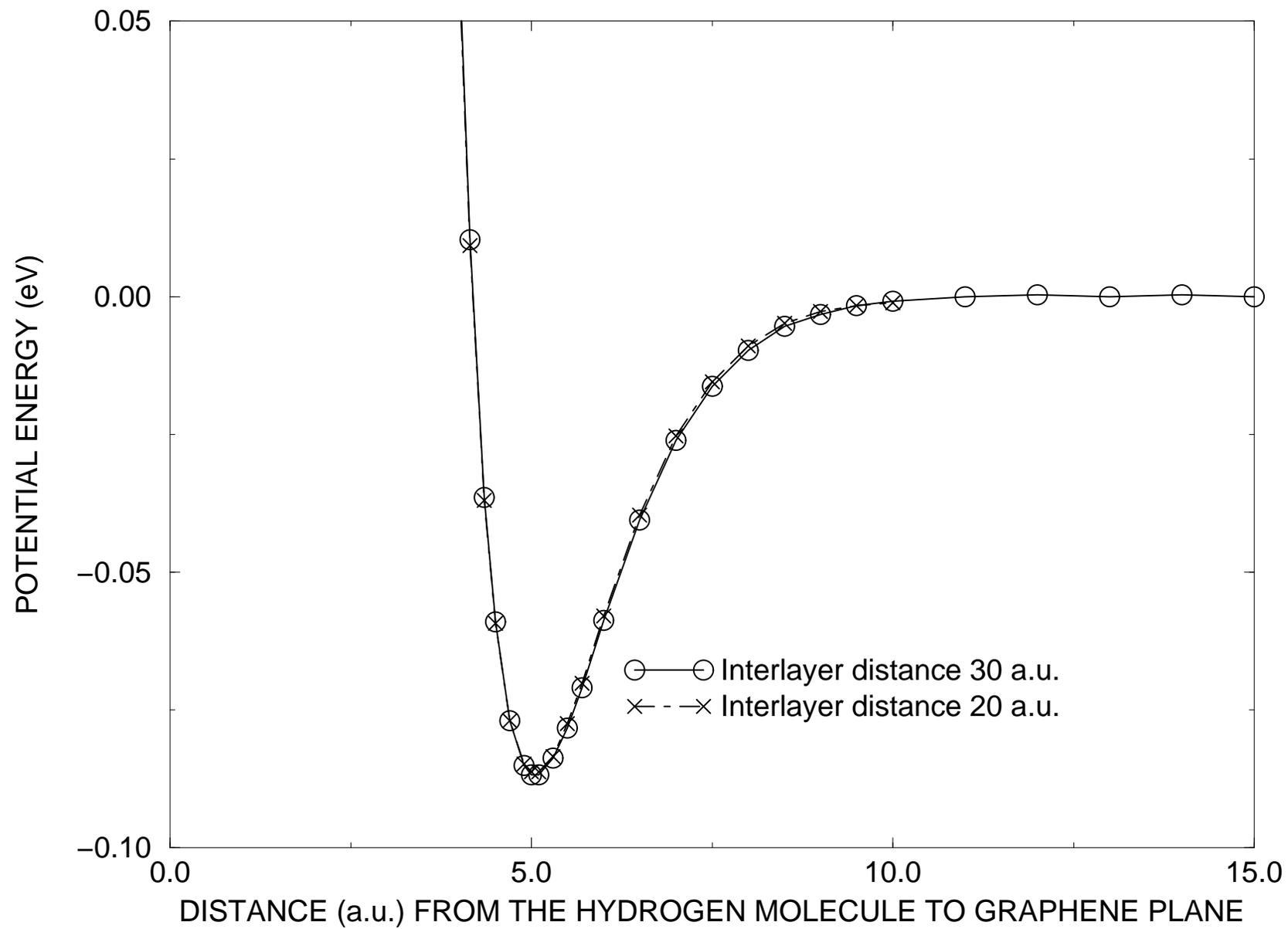
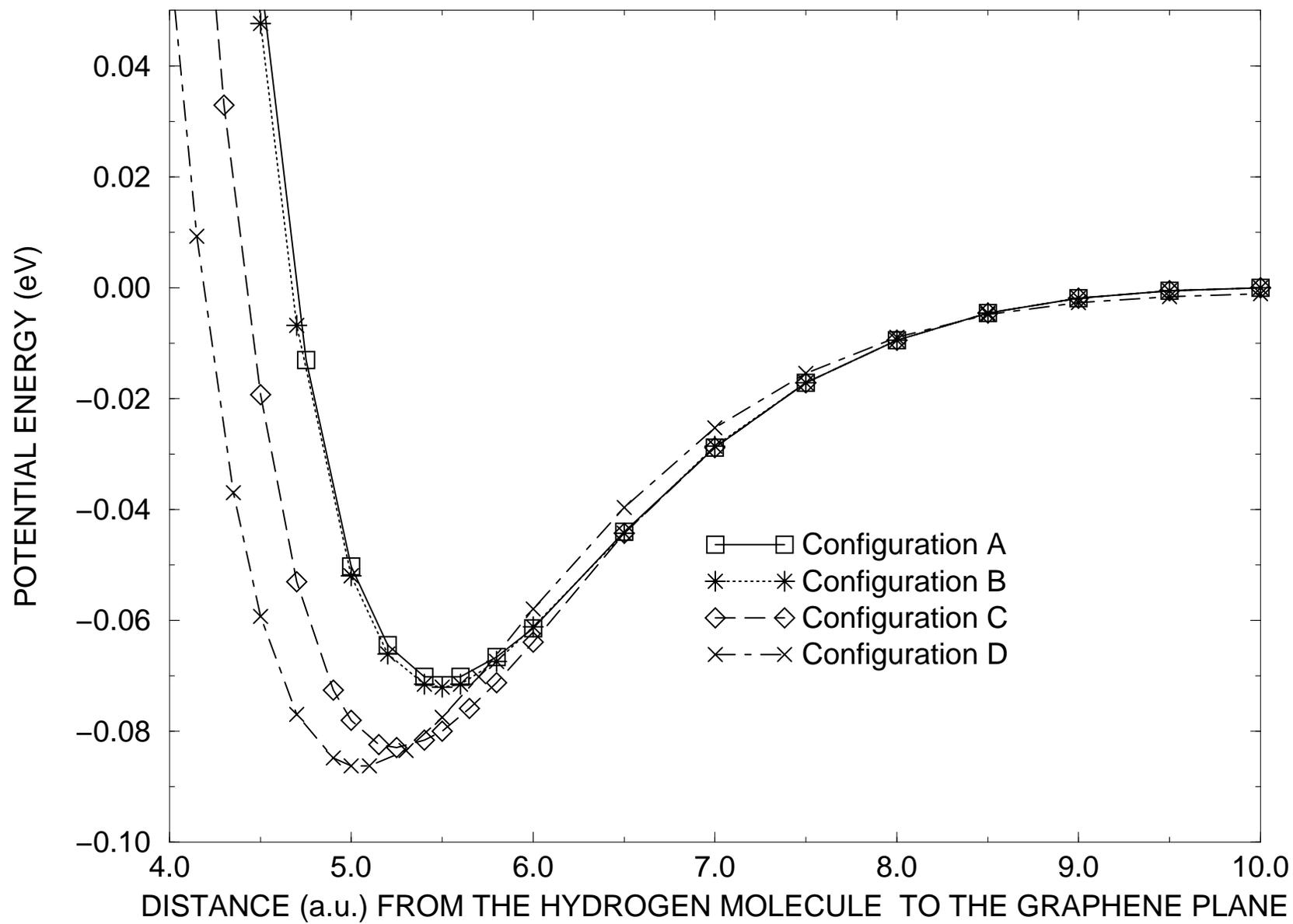
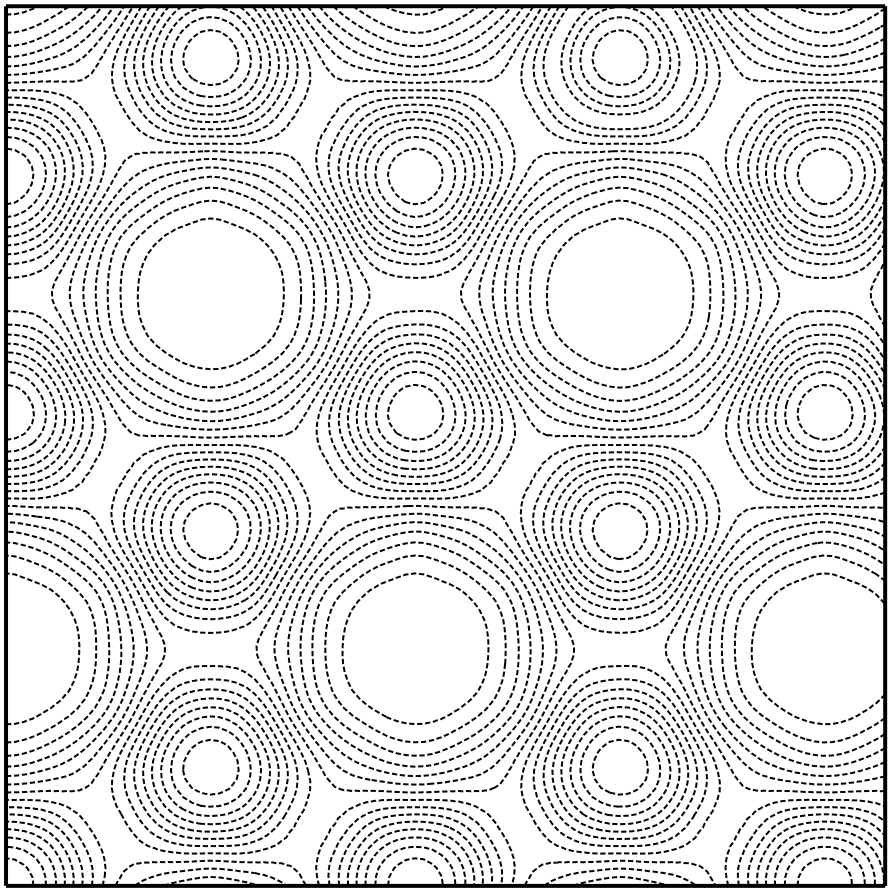
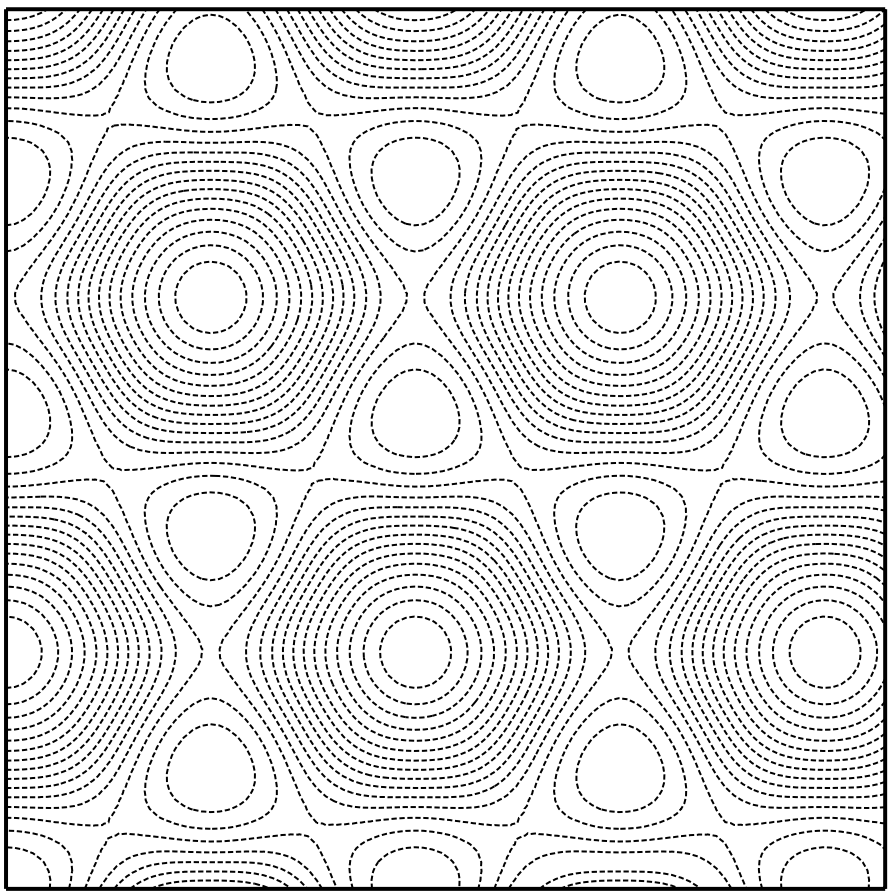
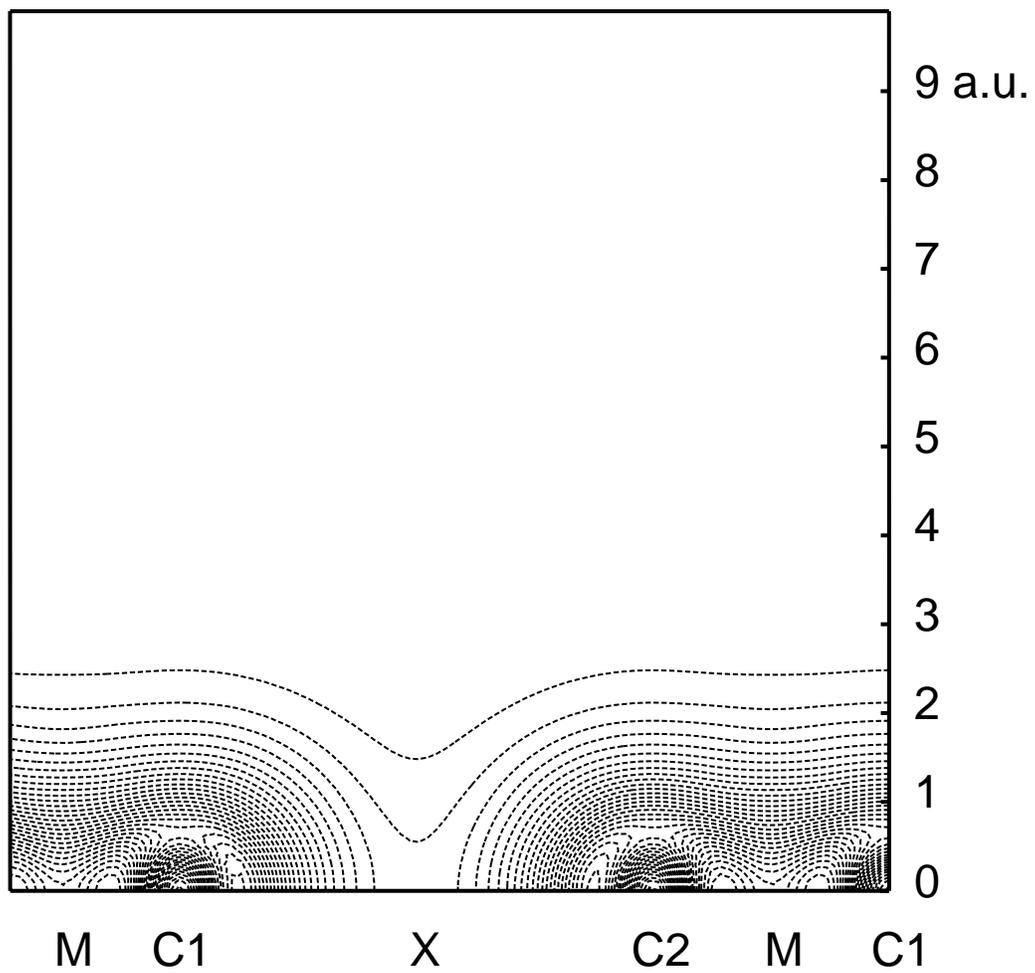
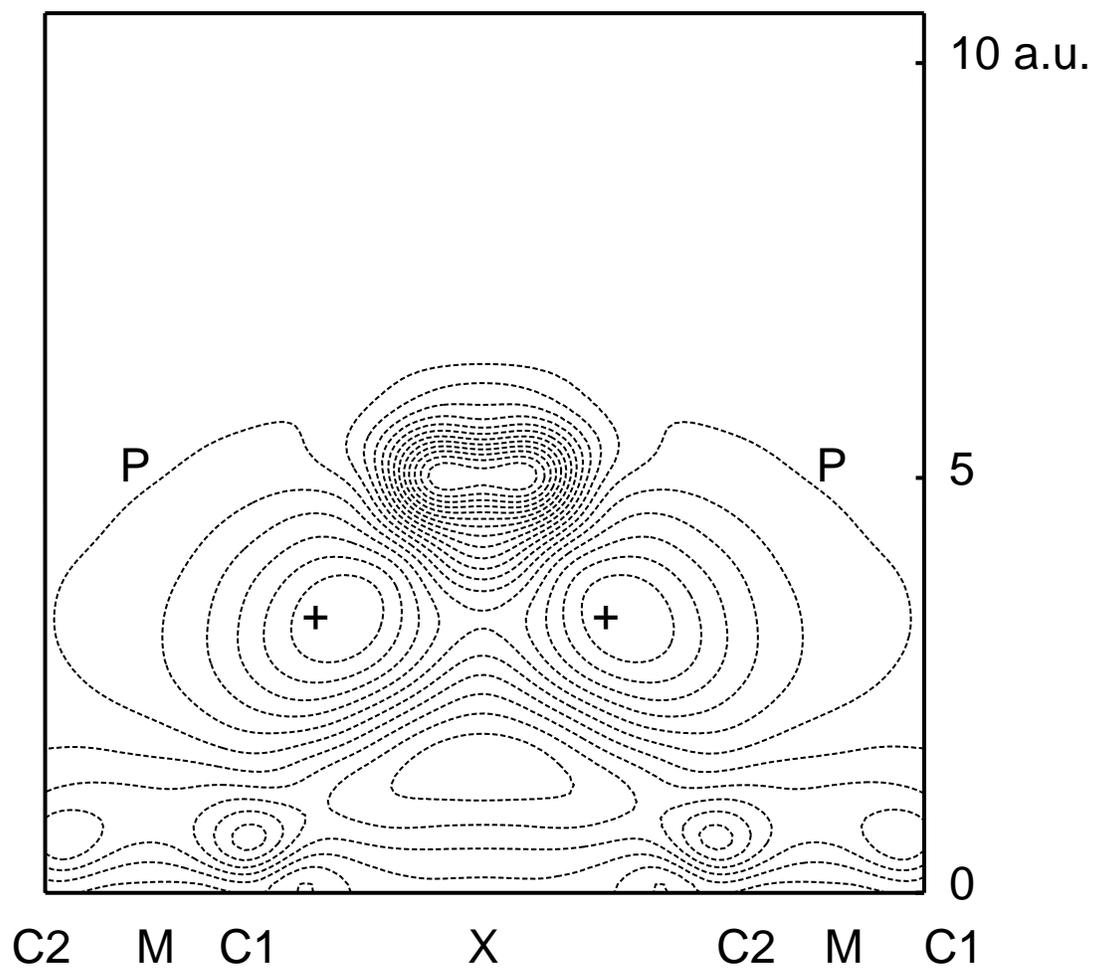
\end{document}